\documentclass[final,3p,times]{elsarticle}

\usepackage{amsthm}

\usepackage{xcolor}
\usepackage{graphicx}
\usepackage{caption}
\usepackage{subcaption}
\usepackage{natbib}
\usepackage{enumitem}
\usepackage{appendix}
\usepackage[section]{placeins}

\setcitestyle{authoryear,open={(},close={)}}

\usepackage{lscape}
\usepackage{algorithm,algorithmic}
\usepackage{lineno}
\usepackage{amsmath}

\journal{Transportation Research Part C}

\begin{document}
\title{Impact of Acceleration/deceleration limits on the String Stability of Adaptive Cruise Control}

\author[1]{Hao Zhou}
\author[1]{Anye Zhou}
\author[2]{Tienan Li}
\author[2]{Danjue Chen}
\author[1]{Srinivas Peeta}
\author[1]{Jorge Laval\corref{cor1}}
\ead{jorge.laval@ce.gatech.edu}
\cortext[cor1]{Corresponding author}
\fntext[label1]{790 Atlantic Dr NW, Atlanta, GA, 30313}
\address[1]{School of Civil and Environmental Engineering, Georgia Institute of Technology, Atlanta, United States}
\address[2]{School of Civil and Environmental Engineering, University of Massachusetts Lowell
, Lowell, United States}

\begin{abstract}
This paper demonstrates that the acceleration/deceleration limits in ACC systems can make a string stable ACC  amplify the speed perturbation in natural driving.
It is shown that the constrained acceleration/deceleration of the following ACCs are likely to cause speed overshoot to compensate for an extra large/small spacing. 
Additionally, we find that the constrained deceleration limits can also jeopardize safety, as the limited braking power produces extra small spacing or even crashes. The findings are validated through experiments on real cars.
The paper suggests that the ACC parameter space should be extended to include the acceleration/deceleration limits considering their significant role exposed here. Through numerical simulations of ACC platoons, we show: i) a marginal string stable ACC is   preferable due to the smaller total queue length and the shorter duration in congestion; ii) congestion waves in a mixed ACC platoon largely depend on the sequence of vehicles provided different acceleration/deceleration limits, and iii) the safety hazard caused by the constrained deceleration limits is more significant in mixed ACC platoons when string unstable ACCs exist.  
\end{abstract}


\maketitle

\section{Introduction}
Adaptive cruise control (ACC) systems have been widely equipped on commercial vehicles to alleviate driving fatigue and enhance safety. The ACC system consists of an upper-level planner and a low-level controller. When ACC is activated, the upper-planner receives the updated information from onboard sensors and plans for the optimal trajectory (e.g., speed, acceleration), and then the low-level controller executes the trajectory by sending low-level commands (e.g., gas/brake) to the car control interface. Recent studies \citep{gunter2019model, li2021ACC} point out that ACC systems equipped on commercial vehicles are not string stable, which implies vehicles driven by ACC can amplify the speed fluctuations and exacerbate the propagation of stop-and-go waves. This discovery inspires many studies to explore the car-following behavior of commercial vehicles driven by ACC systems. For instance, \cite{gunter2019model, li2020trade, shi2021empirical} calibrate parsimonious models (e.g., car-following models, linear feedback controllers) to capture the behavior of ACC systems. The data-driven approach based on a parsimonious model can present a holistic description of an ACC system, however, this approach fails to explain some critical characteristics of an ACC system, such as the effects of upper-level planner and low-level controller, the interactions between them, the build-in desired spacing policy, and the acceleration/deceleration limits. \cite{Zhou2021SignificanceOL} explores the impact of the low-level controller on string stability (SS), however, the acceleration/deceleration limits embedded in the low-level controller are not explicitly analyzed.

Based on the best of our knowledge, few studies have explored the impact of acceleration/deceleration limits, which can significantly jeopardize the SS and safety performance of the ACC system. Moreover, there is also a lack of clarity between SS and speed dampening.

In this study, we first articulate the relationship between SS and speed dampening in situations with or without acceleration/deceleration limits. For traffic engineers, the dampening of speed fluctuations along a string of vehicles is favorable as it alleviates the propagation of stop-and-go waves. Then, a natural question arises as follows. While SS condition borrowed from control theory is defined only given a "small" perturbation, does it necessarily guarantee to dampen speed fluctuations in natural drives? The short answer is 'no'. 
This can be problematic for designing new ACC or CACC algorithms, as most of them rely on the SS condition borrowed from control theory. However, the property of SS may not be sufficient for dampening speed fluctuations. While the SS is only for a small perturbation and seems able to explain "phantom traffic congestion" caused by an unexpected small speed change, the true speed changes in traffic flow are much more complex. It is easy to observe more frequent and large fluctuations or waves in natural driving. For example, a platoon lead might experience up-and-down speed changes in a local road with many curves or grades. Of course, we expect the same "dampening" effect to hold in these scenarios as well. But compared to the "phantom traffic congestion", those more natural but complex speed changes have not drawn enough attention. Unfortunately, so far we don't have any proof or evidence to show the SS is still sufficient for those complex speed-changing cases.   

In the previous study \citep{Zhou2021SignificanceOL}, we introduce the algorithms of an open-source factory ACC system and found that the upper-level planners are already string stable, whereas the low-level controllers play a significant role in the string stability. In that work, we only investigated a homogeneous platoon of ACC-equipped vehicles, and all the perturbations are caused by the ACC systems as well, i.e. the pattern and size of the lead vehicle speed changes are generated by the cruise control system of the lead ACC vehicle. However, in the real-world scenario where vehicles have different acceleration/deceleration limits, the differences in vehicle acceleration/deceleration ability significantly influence the SS of a platoon. 

More importantly, even if the ACC system satisfies SS and speed dampening conditions, the existence of acceleration/deceleration limits can undermine the SS or speed dampening performance. When the perturbation induces the lead vehicle to accelerate/decelerate at a rate beyond the acceleration/deceleration limits of the ego vehicle driven by ACC, the acceleration/deceleration limits will significantly influence the car-following behavior of the ego vehicle. In addition, the perturbation type also matters when acceleration/deceleration limits exist. Specifically, for non-cyclic perturbation (i.e., step-function type speed variations of the lead vehicle), the acceleration/deceleration limits would hinder the vehicle from catching up with the lead vehicle, inducing large spacing. The large spacing then triggers the vehicle to accelerate/decelerate at the limit to follow with the lead vehicle, which inevitably leads to speed overshoot/undershoot and undermines SS. For cyclic perturbation (e.g., sinusoidal speed variations), if the lead vehicle accelerates then decelerates before the ego vehicle (driven by ACC) overshoot (i.e., larger speed fluctuation than the lead vehicle), then SS can be maintained even with the impact of acceleration/deceleration limits. Besides the impact of SS, the acceleration/deceleration limits restrict the vehicle to brake hard when it is entailed for collision avoidance, which jeopardizes the safety performance of the ACC system. This is a critical issue that ACC suppliers should carefully address.

The remainder of the paper is organized as follows. Section 2 articulates the relationship between SS and speed dampening. Section 3 presents the acceleration/deceleration limits in the factory ACC system. Section 4 discusses the impact of acceleration/deceleration limits on string stability and provides relevant empirical evidence. Section 5 articulates the impact of the deceleration limit on safety and provides relevant empirical evidence. Section 6 concludes the paper and points out future directions.



\section{SS of the factory linear ACC and its relationship with dampening}\label{math}

\subsection{SS of the factory linear ACC}
As shown in our previous work \citep{zhou2021impact}, the factory linear ACC module is using the lead vehicle speed $v_{lead}$ from the sensor to plan the desired distance and the target speed/acceleration, which is in contrast with the common design in the literature that adopts the ego speed instead.

Although the small change from $v_{ego}$ to $v_{lead}$ may sound trivial, it could cause huge differences on the SS a linear CF model. Here we investigate the SS of the factory linear model using the transfer function adopted in similar studies.

The dynamics of the factory linear ACC that incorporates $v_{lead}$ is:
\begin{equation}
    v_{ego}(t+\Delta t) = k_v (x_{lead}(t)-x_{ego}(t)-v_{lead}(t) \cdot \tau -\delta) +v_{lead}(t)
    \label{factory-ACC-vlead}
\end{equation}
where $\delta$ is the minimum safe gap between vehicles.

Let $y_{ego}$ and $u_{ego}$ denote the headway and speed variation of the follower vehicle that deviates from the equilibrium state.
\begin{align}
    \dot{y}_{ego} &= \dot{x_{lead}} - \dot{x_{ego}} = (v_{lead}-v_{eq}) - (v_{ego}-v_{eq})= u_{lead} - u_{ego}  \label{spacing-change}    \\
    \dot{u_{ego}} &= \dot{v_{ego}} = k_v \dot{y}_{ego} + (1-k_v \tau) \dot{u_{lead}} \label{speed-change} \\
    & = k_v (v_{lead}-v_{ego}) + (1-k_v \tau) a_{lead}
\end{align}

Note that \eqref{speed-change} is directly from \eqref{factory-ACC-vlead}, which does not rely on the deviation from the equilibrium state. It is the \eqref{spacing-change} that requires the leader and follower to have an initial equilibrium state or just an identical speed. 

Take the Laplace transform and let $U(s) = \mathcal{L}\{u(t)\}$, and $Y(s) = \mathcal{L}\{y(t)\}$.
\begin{align}
    Y(s) &= \mathcal{L}\{ \int (u_{lead}-u_{ego}) dt \} = \frac{1}{s} [\mathcal{L}\{ u_{lead}(t)\} -\mathcal{L}\{ u_{ego}(t) \}]= \frac{1}{s} [U_{lead}(s) - U_{ego}(s)]   \label{Y(s)}   \\
    sU_{ego}(s) &= k_v Y(s) +(1-k_v \tau) \cdot s U_{lead}(s) \label{Un(s)}
\end{align}

Substitute \eqref{Y(s)} into \eqref{Un(s)}, we have the transfer function:
\begin{align}
    sU_{ego}(s) &=  \frac{k_v}{s} [U_{lead}(s) - U_{ego}(s)] + (1-k_v \tau) s U_{lead}(s) \nonumber  \\
   \Gamma(s) &= \frac{U_{ego}(s)}{U_{lead}(s)} = \frac{(1-k_v \tau)s^2+k_v}{s^2+k_v}
   \label{transfer-function}
\end{align}
where we need $|1-k_v\tau|<1$ to guarantee the SS. Notice that $k_v>0$ and vary at different speed levels. Time headway is usually a constant with short, medium, and far options. Thus the SS condition reduces to:
\begin{equation}
    k_v (v) < 2/\tau
    \label{SS-condition}
\end{equation}

The above condition \eqref{SS-condition} indicates smaller $k_v$ helps SS. Recall that $k_v$ can be understood as the sensitivity of the CF model to the deviation from the desired gap, it determines how fast the ACC tries to move towards the desired gap. 

Besides the traditional transfer function method from the control theory, the author also shows an ODE-based solution can find the same SS condition in \eqref{SS-condition}; see Appendix.





\subsection{Relationship between the SS and speed dampening without acceleration limit}
First, we release the constraint of acceleration limit and investigate whether a 'string stable' model given a 'small' perturbation can actually guarantee the 'speed dampening' in natural driving featuring complex 'up-and-down speed trajectories. 

The transfer function has been widely studied in the literature, however, to the best of our knowledge, all of these studies assume an initial equilibrium state with a "small" perturbation like a single sine wave. The relation between the SS and dampening in real driving is left unknown. 

Now we prove the SS is a sufficient condition for speed dampening with infinite acceleration/deceleration bounds.


\begin{proof}
A natural driving leader speed profile can be decomposed as a sum of sinusoidal waves according to Fourier theory: 
\begin{equation}
    v_{lead}(t) = M_1 sin(\omega_1 t) + M_2 sin(\omega_2 t) +...+M_n sin(\omega_n t)
\end{equation}
where $M_i$ and $\omega_i$ denote the magnitude and frequency of the $i$ component in the speed perturbation. 

Apply Laplace transform which has the additive property:
\begin{equation}
     \mathcal{L}\{v_{lead}(t)\} =  \mathcal{L}\{M_1 sin(\omega_1 t) \} +  \mathcal{L}\{M_2 sin(\omega_2 t) \}+... + \mathcal{L}\{M_n sin(\omega_n t)\}
\end{equation}

Thus the input signal $U_s$ can be decomposed in the s-domain:
\begin{equation}
     U(s) = U_1(s) +  U_2(s)+ ... U_n(s)
\end{equation}
where $U_i(s) = \mathcal{L}\{M_i sin(\omega_i t) \}$ and $U(s) = \mathcal{L}\{v_{lead}(t)\}$.

Denote the output signal $Y(s) = \mathcal{L}\{v_{ego}(t)\}$.

For a parallel system, the transfer function has the additive property, which gives:
\begin{align}
     \frac{Y(s)}{U(s)} &= \frac{\Gamma_1(s)U_1 (s) + \Gamma_2(s) U_2 (s) + ... \Gamma_n(s) U_n (s)}{ U_1 (s) +  U_2 (s) + ... U_n (s)} \\
     &= \Gamma(s) \leq 1 \quad \textit{if} \quad  \Gamma_i(s) \leq 1 \quad \forall{i}
\end{align}
\end{proof}

If the follower is an LTI system, we have $\Gamma_i=\Gamma$, $\forall{i}$.

\begin{figure}[htbp]
\centering
\begin{subfigure}[t]{0.35\linewidth}
    \includegraphics[width=\linewidth]{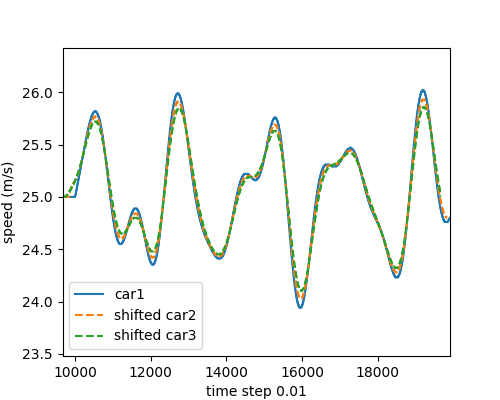}
    \caption{speed profile}
\end{subfigure}%
\begin{subfigure}[t]{0.35\linewidth}
    \includegraphics[width=\linewidth]{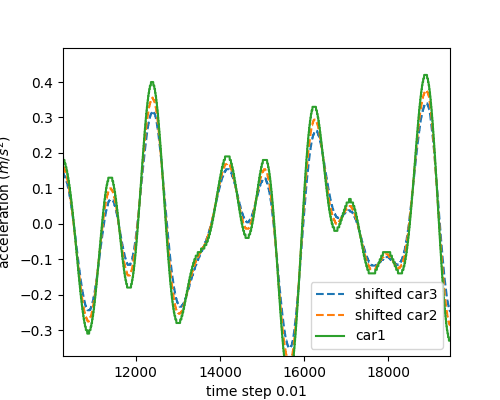}
    \caption{acceleration profile}
\end{subfigure}%
    \caption{Speed and acceleration of a platoon that dampens using the factory linear ACC}
    \label{speed-accel-dampening}
\end{figure}

To illustrate such a relationship between SS and speed dampening, Fig.\ref{speed-accel-dampening} shows the speed and acceleration of a three-vehicle platoon that dampens the speed changes which consists of several sinusoidal waves. The lead (car1) trajectory in (a) is to simulate the natural driving with small, occasional, and cyclic perturbations.  From (b) it is clear that the follower acceleration has a smaller amplitude in all cycles, which leads to the dampening speeds in (a). One can easily show the acceleration dampens because the acceleration to acceleration transfer function is the same as the speed to speed transfer function according to the property of Laplace transformation.

\FloatBarrier


\section{Acceleration/deceleration limits in ACCs}
In the last section, we have shown the SS condition of the linear factory ACC and prove that SS can guarantee speed dampening given infinite acceleration/deceleration bounds. However, infinite limits are obviously not realistic in market ACC systems. To study the realistic acceleration/deceleration limit values and their impact, we first investigate the design of acceleration/deceleration bounds in the open-source ACC platform, Openpilot. Empirical data from recent market ACC vehicles are also shown for more evidence.


\subsection{Design of the acceleration/deceleration limits in Openpilot}

\begin{figure}[htbp]
\centering
\begin{subfigure}[t]{0.45\linewidth}
    \includegraphics[width=\linewidth]{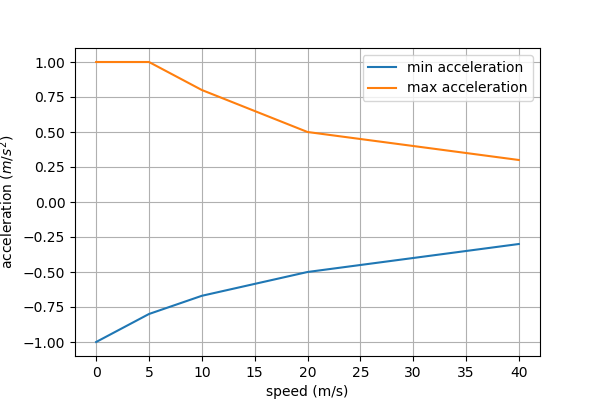}
    \caption{Linear ACC}
\end{subfigure}
\begin{subfigure}[t]{0.45\linewidth}
    \includegraphics[width=\linewidth]{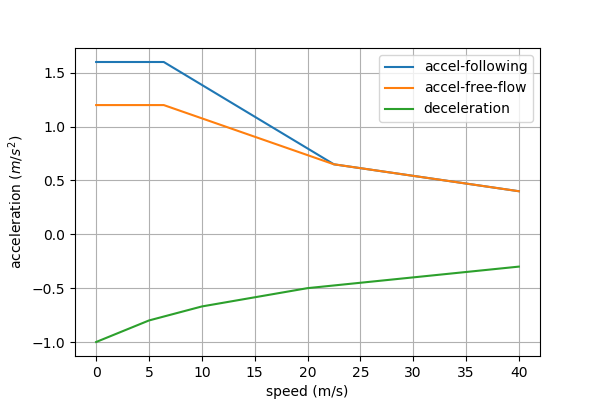}
    \caption{MPC planner}
\end{subfigure}
\caption{Acceleration/deceleration limits in Openpilot}
\label{OP-limits}
\end{figure}

Now we show the acceleration/deceleration limits for both the linear and MPC planner in Openpilot; see Fig.\ref{OP-limits}. The acceleration limits in both the linear planner and the MPC are piece-wise linear, similar to the linear acceleration model. Similarly, we see a linear deceleration model, which still needs more justification. Also, note that the MPC planner shows two different modes for the acceleration limits, where a more conservative acceleration bound is used when there is no CF state.

Recall the linear acceleration model of human drivers \cite{Laval2014APM}:
\begin{align}
    \frac{dv}{dt} &= (v_c -v)\beta 
    \label{linear_accel_model}
\end{align}

Correspondingly, we assume the acceleration limit also has a linear shape:
\begin{align}
    a^*(v) & = a_0 + (v_c -v)\beta
    \label{accel_bound}
\end{align}
where the empirical data suggest that $a_0 \approx 0.4 $, $v_c \approx 40 m/s $ and $\beta \approx 0.015$.

For a vehicle that accelerates using the maximum acceleration in \eqref{accel_bound} with a initial speed $v_0$, its speed can be obtained as:
\begin{align}
    v_{ego}(t) &= (v_0-v_c-a_0/\beta)e^{-\beta t} + a_0/\beta + v_c 
    \label{max_speed_traj}
\end{align}

\subsection{Acceleration/deceleration data from market ACCs}
\begin{figure}[htbp]
\centering
\begin{subfigure}[t]{0.33\linewidth}
    \includegraphics[width=\linewidth]{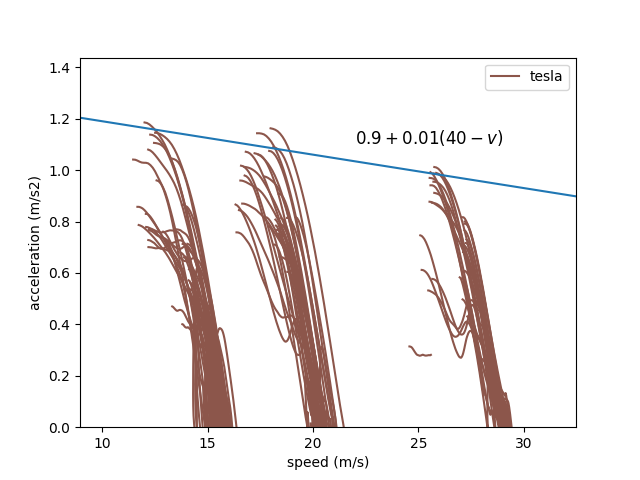}
    \caption{Tesla Model X}
\end{subfigure}%
\begin{subfigure}[t]{0.33\linewidth}
    \includegraphics[width=\linewidth]{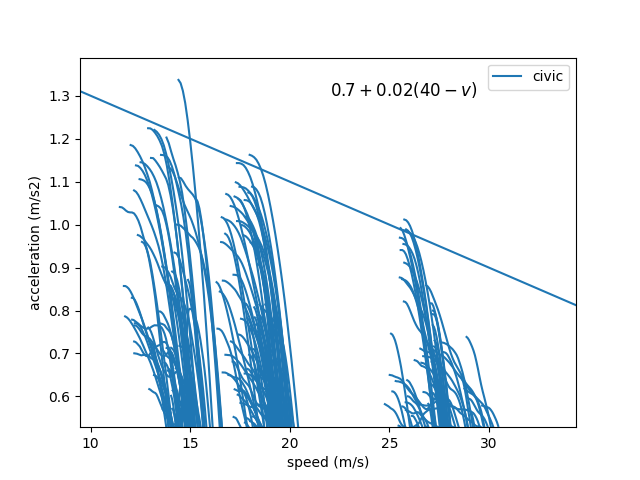}
    \caption{Honda Civic}
\end{subfigure}%
\begin{subfigure}[t]{0.33\linewidth}
    \includegraphics[width=\linewidth]{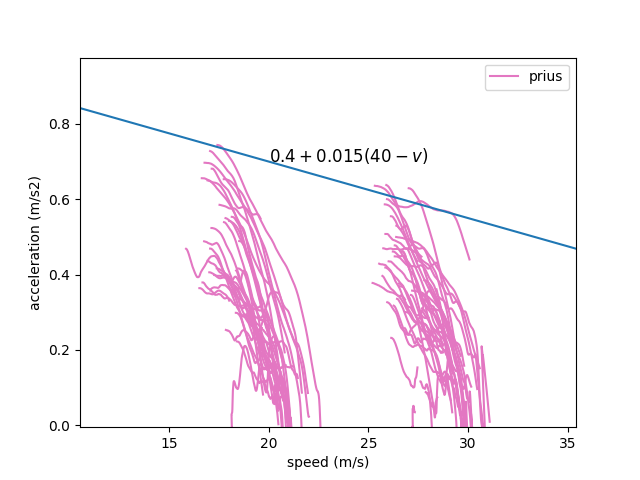}
    \caption{Toyota Prius}
\end{subfigure}%
\caption{Acceleration limits in market ACC vehicles}
\label{real-acceleration-limits}
\end{figure}

We conjecture that market ACC systems adopt a similar pattern but different values of the acceleration/deceleration limits in Openpilot. To provide more evidence, now we show the empirical data of the acceleration/deceleration values in real drives that collected three recent ACC car models, including a Tesla Model X, a Honda Civic, and an electric car Toyota Prius. The drives are designed as oscillations created by a human-driven leader, who decelerated, remained its speed for a few seconds, and then accelerated to revert its origin speed level.  More experiment details can be seen in \citep{li2021ACC},

In each drive, the acceleration process is extracted and shown as a single curve Fig.\ref{real-acceleration-limits}. Notice that the human driver lead vehicle adopts larger acceleration/deceleration values in those experiments, which create extra small or large gaps for the ACC vehicle to compensate for. 
Hence, we believe the tipping points at the curves in Fig.\ref{real-acceleration-limits} can approximately show the acceleration limits in the ACC design since they correspond to much larger or smaller gaps compared to equilibrium. We also connect the tipping points at different speed levels, which leads to approximately linear acceleration limits as similar to Openpilot Fig.\ref{OP-limits}, which also provides more evidence to our theory in \eqref{accel_bound}. 

\FloatBarrier

\begin{figure}[htbp]
\centering
\begin{subfigure}[t]{0.33\linewidth}
    \includegraphics[width=\linewidth]{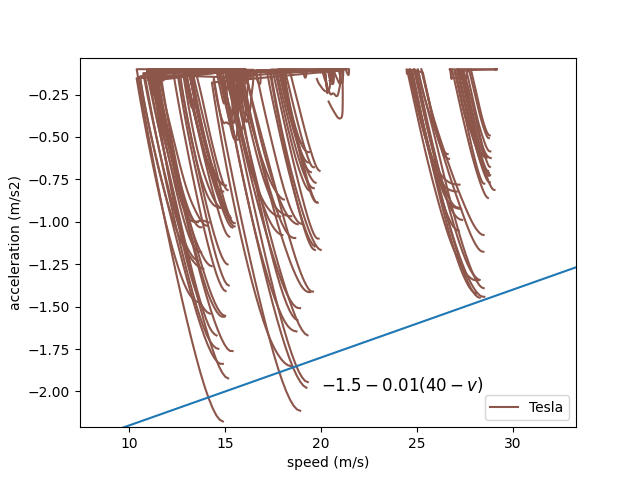}
    \caption{Tesla Model X}
\end{subfigure}%
\begin{subfigure}[t]{0.33\linewidth}
    \includegraphics[width=\linewidth]{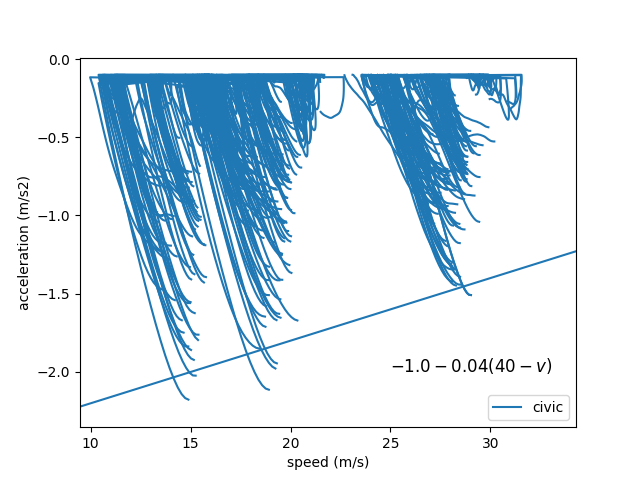}
    \caption{Honda Civic}
\end{subfigure}%
\begin{subfigure}[t]{0.33\linewidth}
    \includegraphics[width=\linewidth]{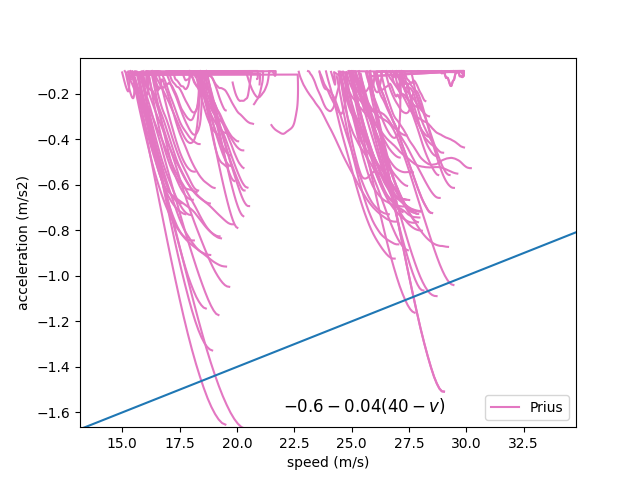}
    \caption{Toyota Prius}
\end{subfigure}%
\caption{Deceleration limits in commercial ACC vehicles:}
\label{real-deceleration-limits}
\end{figure}

Similarly, we see the deceleration bounds in market ACCs in Fig.\ref{real-deceleration-limits}. The tipping point of each drive indicates the ACC deceleration limit at that speed level, and connecting them also produces a linear deceleration bound. However, the authors argue that such a linear deceleration design is not as physically straightforward as the acceleration case constrained by engine power. We conjecture a smaller deceleration rate at higher speeds is designed for driving comfort, which may pose a safety issue and needs more justification. 

\FloatBarrier







\section{Impact of acceleration/deceleration limits on traffic operation}

Now we study the impact of those acceleration/deceleration limits found in factory ACCs. To study the impact of SS or speed dampening, we assume the lead vehicle is a high-performance car that produces an abrupt speed change, either cyclic or non-cyclic depending on whether it reverts to its origin speed.

\subsection{Non-cyclic lead speed perturbation}

\begin{figure}[htbp]
\centering
\begin{subfigure}[t]{0.49\linewidth}
    \includegraphics[width=\linewidth]{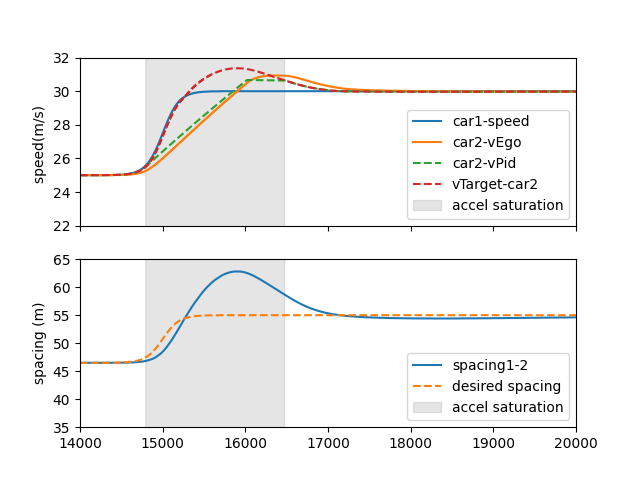}
    \caption{speed and spacing of the two vehicles}
\end{subfigure}
\begin{subfigure}[t]{0.49\linewidth}
    \includegraphics[width=\linewidth]{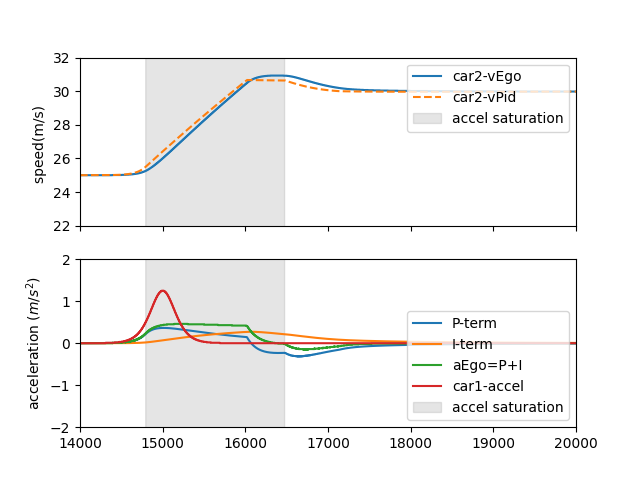}
    \caption{low-level details of the follower}
\end{subfigure}%
\caption{The impact of acceleration limit with a non-cyclic perturbation: car1 is the human-driven leader and car2 is the following ACC vehicle equipped with a linear ACC planner and PI low-level controller. The shaded area indicates the acceleration limit is reached.}
\label{non-cyclic-impact}
\end{figure}

For the non-cyclic speed perturbation, assume the lead vehicle increases/decreases its speed from $v_{lead}(T_0)$ at time $T_0$ to its final speed $v_{lead}(T_5)$ at time $T_5$. The speed trajectory of the leader can be seen as the blue curve in Fig.\ref{non-cyclic-impact} (a). To show the impact of ACC acceleration limit, we assume the lead vehicle (car1) is a performance car or driven by an aggressive driver who presses the gas pedal hard, which can produce a larger acceleration compared to a normal vehicle; see the acceleration (red curve) in Fig.\ref{non-cyclic-impact} (b).

\FloatBarrier

Now we investigate the responses of the following ACC vehicle and study the impact of acceleration limit on speed overshoot. 

In Fig.\ref{non-cyclic-impact} (a), from $T_0$ to $T_1$, one can see the lead vehicle accelerates faster and has a speed advantage over the follower ACC car, which makes the spacing keep growing to its maximum at time $T_1$; see the blue curve in the lower graph of Fig.\ref{non-cyclic-impact} (a). Let $s(t)$ denote the spacing at time $t$, we find that the spacing between $T_0$ and $T_1$ has increased to an exceedingly large value compared to the desired spacing $s(T_5) = \tau \cdot v_{lead}(T_5)+\delta$. We argue that, the extra spacing is the main cause that leads the follower ACC to overshoot, i.e. $v_{ego}(t)>v_{lead}(T_1)$ after $T_1$. 

To illustrate this finding, we start to examine the planning and low-level control process of the following ACC vehicle. According to the planner design, the $v_{target} = v_{lead} + k(s-\tau v_{lead}-\delta) $. Obviously, starting from $T_1$, the true spacing is larger than the desired spacing, which leads to $v_{target}(t)>v_{lead}(t)$ for $T_1 \le t \le T_5$. Also note that the planning speed $v_{target}$ directly results from the spacing, which does not account for the acceleration limits. Instead, it is the low-level controller that actually factors the acceleration limit and move the true speed smoothly. Specifically, $v_{\text{pid}} + a_{max} \cdot dt $ if $v_{\text{target}}>v_{\text{pid}}+a_{max} \cdot dt$.


Now we further investigate how $v_{pid}$ incorporates the acceleration limit and changes the true speed of the follower. From $T_0$ to $T_1$, the low-level setpoint $v_{pid}$ updates itself using $v_{pid} =v_{pid} + a_{max} dt$, which means it is moving at the designed acceleration limit of the ACC. This condition holds until $v_{target}$ drops to the same value of $v_{pid}$.


From $T_2$, the low-level setpoint starts to drop, but the true speed $v_{ego}$ has experienced more overshooting due to the integral accumulation; see the orange line in the lower part of Fig.\ref{non-cyclic-impact} (b). The overshoot value caused by the integral term also depends on the accumulative tracking error of the specific low-level controller. The detailed impact of a fast/slow low-level controller can be seen in \cite{Zhou2021SignificanceOL}.

From $T_4$ to $T_5$, the low-level controller is no longer subject to acceleration or deceleration constraint, and $v_{pid}$ is perfectly led by the $v_{target}$ to close the extra gap. 

To summarize, we see the acceleration limit in ACC design has prevented the low-level setpoint from tracking the planning target $v_{target}$. This causes the true spacing to keep increasing to an exceeding value compared to the desired spacing of ACC. During the process of catching up with the leader, the ACC follower is accelerating using the designed acceleration limit, or \eqref{accel_bound} equivalently. The extra gap stops increasing when the ACC vehicle reaches and overshoots  $v_{lead}$. Then the follower starts to close the gap. 

We show that the maximum overshoot speed of the follower ACC is solvable, provided the lead vehicle trajectory. The detailed process can be seen in Appendix.

\begin{figure}[htbp]
\centering
\begin{subfigure}[t]{0.48\linewidth}
    \includegraphics[width=\linewidth]{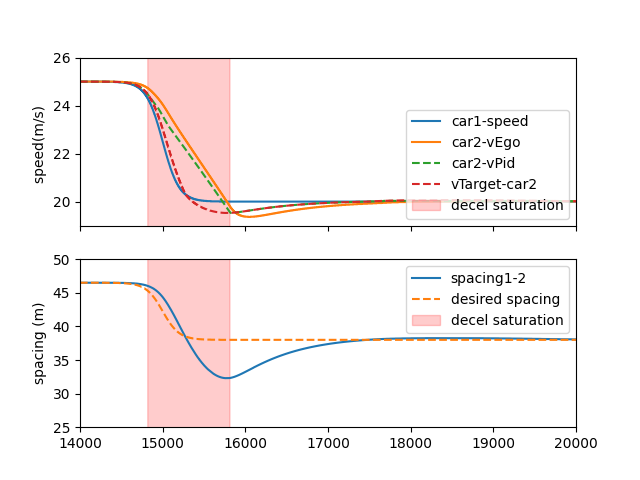}
    \caption{speed and spacing of the two vehicles}
\end{subfigure}
\begin{subfigure}[t]{0.48\linewidth}
    \includegraphics[width=\linewidth]{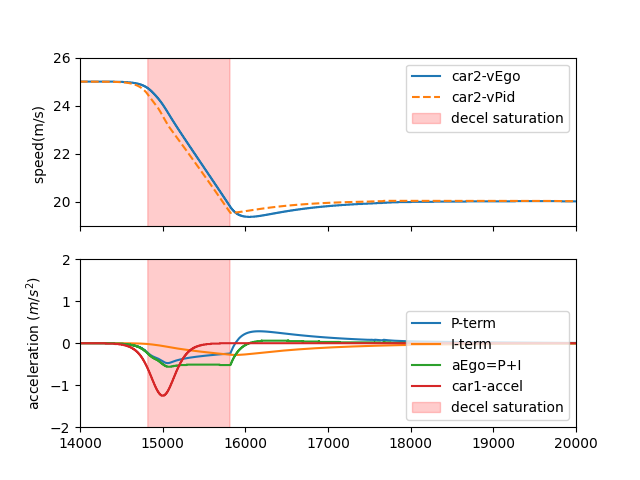}
    \caption{low-level details of the follower}
\end{subfigure}%
\caption{The impact of deceleration limit with a non-cyclic perturbation: the shaded area indicates the deceleration limits are being used.}
\label{non-cyclic-impact-deceleration}
\end{figure}

From the above analysis, one can see the acceleration limit can ruin the dampening condition if the follower cannot close the extra spacing before its speed equals the leader speed. Similarly, for the deceleration case, the deceleration limit would undermine the dampening condition if the follower cannot brake enough to the equilibrium spacing $\tau v_{lead}^1+\delta$; see Fig.\ref{non-cyclic-impact-deceleration}.

\FloatBarrier

\subsection{Cyclic lead speed perturbation}

The cyclic lead speed change is more likely to alleviate the negative impact on speed dampening caused by acceleration/deceleration limits because the lead vehicle helps to close the extra gap by reverting to its origin speed when the follower also tries to catches up using its constrained acceleration/deceleration. 

\begin{figure}[htbp]
\centering
\begin{subfigure}[t]{0.48\linewidth}
    \includegraphics[width=\linewidth]{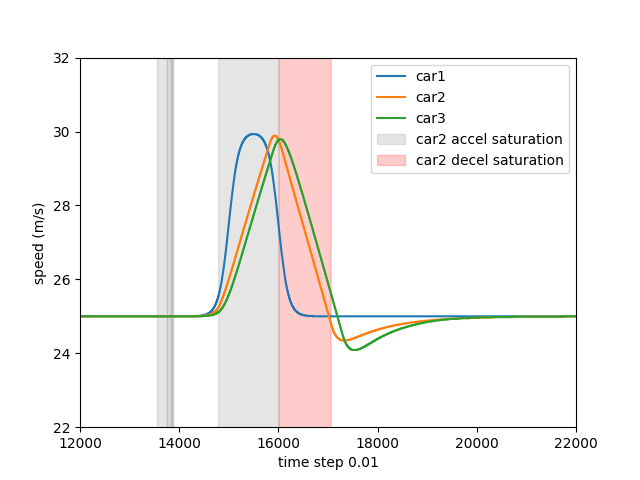}
    \caption{accelerate then decelerate}
\end{subfigure}
\begin{subfigure}[t]{0.48\linewidth}
    \includegraphics[width=\linewidth]{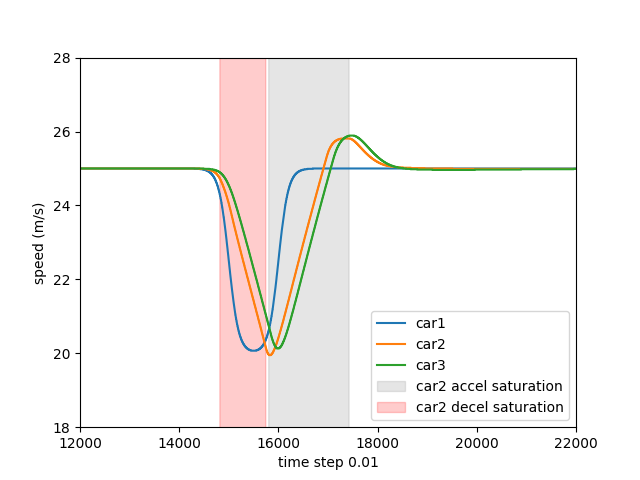}
    \caption{decelerate then accelerate}
\end{subfigure}%
\caption{The impact of acceleration/deceleration limits on platoon speeds with the cyclic perturbations.}
\label{cyclic-platoon}
\end{figure}
\FloatBarrier

We first show the results of a single perturbation in Fig.\ref{cyclic-platoon}, which suggests the SS of the ACC model can be deteriorated by the acceleration/deceleration limits. 

\begin{figure}[htbp]
\centering
\begin{subfigure}[t]{0.48\linewidth}
    \includegraphics[width=\linewidth]{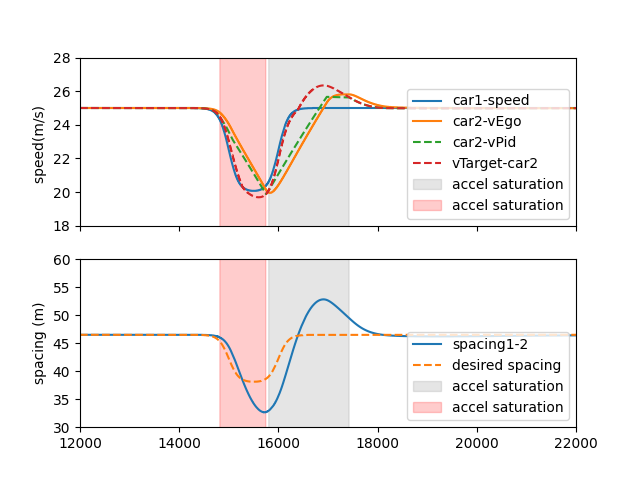}
    \caption{speed and spacing of the two vehicles}
\end{subfigure}
\begin{subfigure}[t]{0.48\linewidth}
    \includegraphics[width=\linewidth]{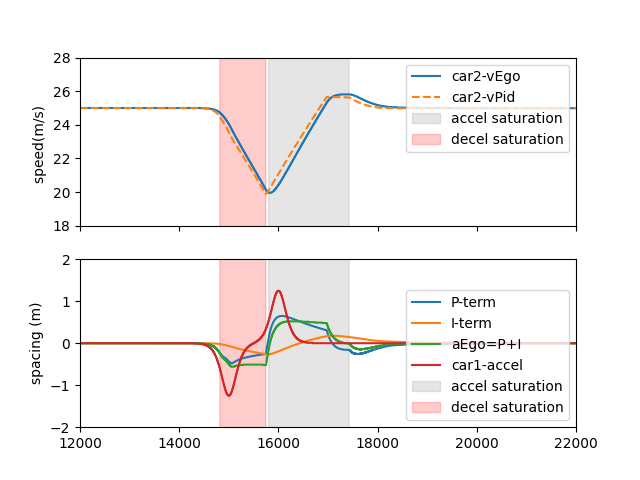}
    \caption{low-level details of the follower}
\end{subfigure}%
\caption{The impact of acceleration/deceleration limit with a cyclic perturbation.}
\label{cyclic-impact}
\end{figure}

In Fig.\ref{cyclic-impact}, we repeat similar experiments as the non-cyclic case, which shows that the ACC follower does not overshoot too much at the peak/valley if the lead vehicle adopts a cyclic speed pattern. We argue that the overshooting at those points are not significant because the lead vehicle happens to change its speed and help reduces the extra small or large gap. However, the overshoot at the end of a wave is still significant, similar to what we have seen from the non-cyclic perturbations. 

\FloatBarrier

\subsection{Empirical evidence}

We have shown that acceleration/deceleration limits in following ACC vehicles can cause super large or small spacings and thus lead to speed overshooting. For validation, we found more evidence in the empirical data from the ACC experiments \citep{li2021ACC}.

\begin{figure}[htbp]
\centering
\begin{subfigure}[t]{0.33\linewidth}
    \includegraphics[width=\linewidth]{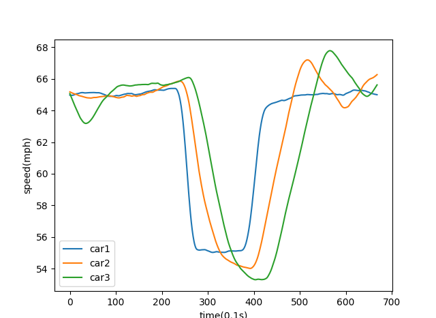}
    \caption{speed}
\end{subfigure}
\begin{subfigure}[t]{0.33\linewidth}
    \includegraphics[width=\linewidth]{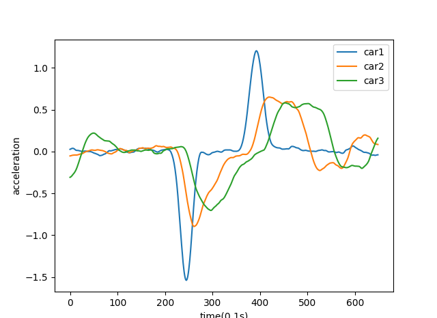}
    \caption{acceleration}
\end{subfigure}%
\begin{subfigure}[t]{0.28\linewidth}
    \includegraphics[width=\linewidth]{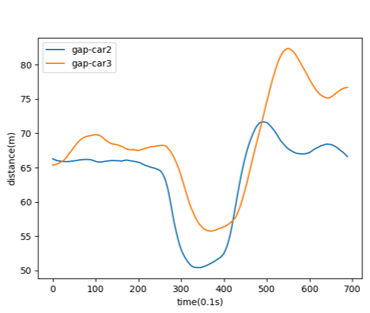}
    \caption{gap}
\end{subfigure}
\caption{Impact of ACC acceleration limit in Civic experiments}
\label{civic-impact}
\end{figure}
\FloatBarrier

\begin{figure}[htbp]
\centering
\begin{subfigure}[t]{0.31\linewidth}
    \includegraphics[width=\linewidth]{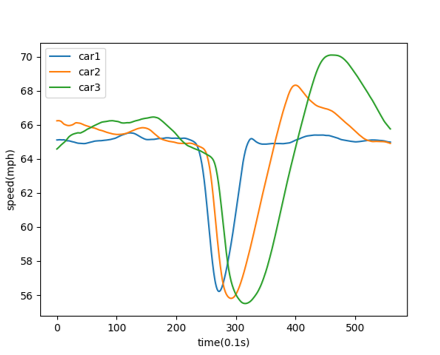}
    \caption{Tesla}
\end{subfigure}
\begin{subfigure}[t]{0.33\linewidth}
    \includegraphics[width=\linewidth]{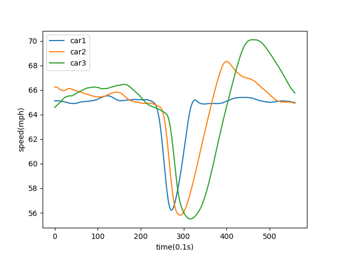}
    \caption{Prius}
\end{subfigure}%
\caption{Impact of ACC acceleration limit in Model X and Prius}
\label{tesla-prius}
\end{figure}

Similar speed overshooting caused by ACC acceleration limits is found on the other two car models (see Fig.\ref{tesla-prius}.
\FloatBarrier



\section{Impact of the deceleration limit on safety}

From Openpilot and the empirical data from other commercial ACCs, we found that the deceleration rate of ACC vehicles is also constrained, especially at high speeds, although the true deceleration capability should not be limited as the acceleration case. While the acceleration limit leads to an extra large spacing, the deceleration constraint leads to the opposite, which can be a super small headway, causing a significant threat to a rear-end collision.

Now we study the abrupt deceleration process mathematically. Assume the lead vehicle decreases it speed due to the congestion ahead, the following ACC vehicle needs to decelerate with a required acceleration rate $a_{required}$ to prevent a collision. For the linear planner, we require the $v_{target}$ to generate a safe deceleration rate to avoid the collision:
\begin{align}
    k(v_{ego}) (v_{lead}(t)-v_{ego}(t)-\tau \cdot a_{lead}(t)) & \le a_{required} \nonumber  \\
    k(v_{ego}) & \ge \frac{a_{required}}{v_{lead}(t)-v_{ego}(t)-\tau \cdot a_{lead}(t)}
    \label{safety-k-condi}
\end{align}

The \eqref{safety-k-condi} requires a lower bound of $k(v)$ to produce enough deceleration to avoid a collision. The \eqref{safety-k-condi} also indicates a smaller time headway $\tau$ requires larger sensitivity.   

Recall the SS condition in \eqref{SS-condition} gives an upper bound of $k(v)$. Combining the two, we obtain:
\begin{align}
    \frac{a_{required}}{v_{lead}(t)-v_{ego}(t)-\tau \cdot a_{lead}(t)} \le k(v) \le \frac{2}{\tau}
\end{align}


While the SS requires the $k$ to be smaller than $2/\tau$, the safety constraint requires a lower bound.

\begin{figure}[htbp]
    \centering
    \includegraphics[width = 0.6\textwidth ]{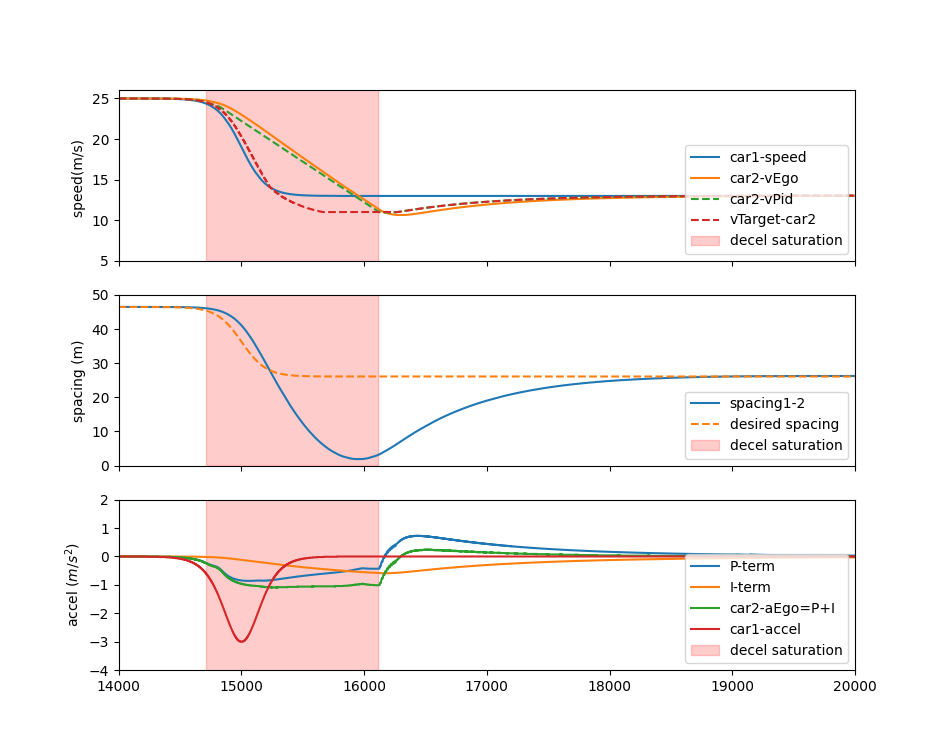}
    \caption{Safety hazard caused by the deceleration limit in a fast congestion wave}
    \label{very-short-distance}
\end{figure}

To simulate the safety hazard, we apply an abrupt deceleration profile for the lead vehicle and investigate the behaviors of the Openpilot ACC follower. The results are shown in Fig.\ref{very-short-distance}, where the lead vehicle decelerates from 25 m/s to half of its original speed, and the follower almost crashes into it with a minimum spacing close to zero.   

\begin{figure}[htbp]
\centering
\begin{subfigure}[t]{0.48\linewidth}
    \includegraphics[width=\linewidth]{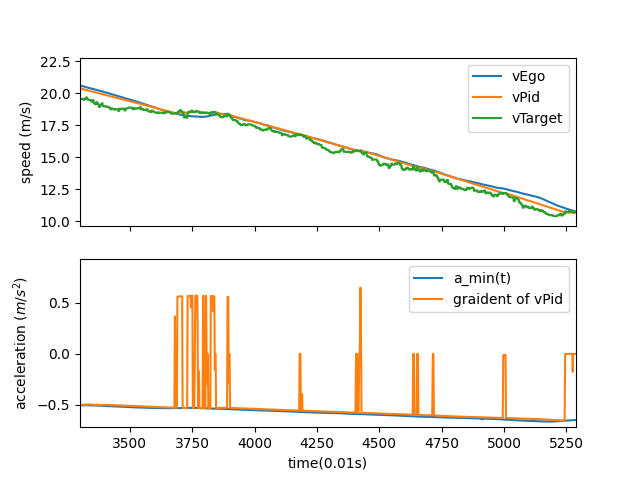}
    \caption{A small deceleration limit leads to the unsafe brake}
\end{subfigure}%
\begin{subfigure}[t]{0.48\linewidth}
    \includegraphics[width=\linewidth]{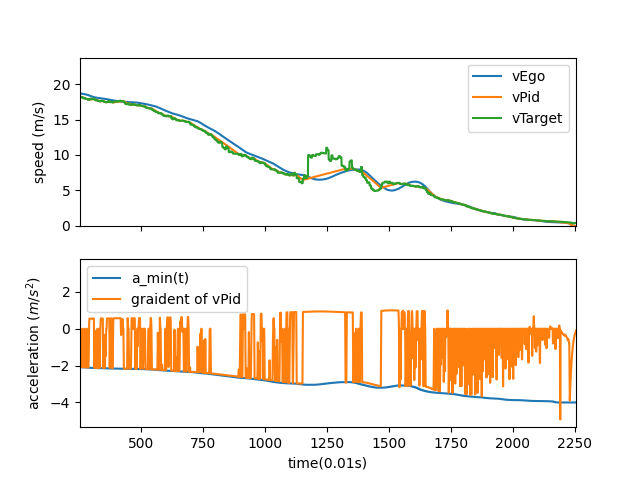}
    \caption{Large deceleration limit guarantees the safe brake}
\end{subfigure}
\caption{Comparing small and large deceleration limits on the safe braking after a stopped car.}
\label{safe-stopping}
\end{figure}

To validate the impact of deceleration bounds on safe braking, we also conduct the experiments using a real car with the same ACC algorithm but a modified deceleration bound from Openpilot. We adopted a small and large deceleration bound and tested the braking process before an intersection where the ACC vehicle needs to safely decelerate to the lead stopped a vehicle at the red lights. The comparison results are shown in Fig.\ref{safe-stopping}. The $a_{min}(t)$ curve shows our design of the deceleration bound, and the gradient of $v_{pid}$ is the derivative of time. The derivatives are jerky, indicating some fluctuations in the planning speed $v_{target}$. However, we clearly see the low-level set-point $v_{pid}$ cannot move slower than the deceleration limit $a_min(t)$ in the spirit of the ACC design. In Fig.\ref{safe-stopping} (b), a large deceleration value is used where the car can successfully brake to speed zero. In contrast, in Fig.\ref{safe-stopping} (a), the more constrained deceleration limit failed to brake in time, where the human driver was forced to take over at speed $10m/s$ when the lead spacing was already unsafe. 

\begin{figure}[ht]
\centering
\begin{subfigure}[t]{0.33\linewidth}
    \includegraphics[width=\linewidth]{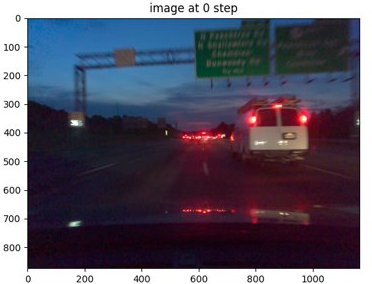}
    \caption{Front view at step 0}
\end{subfigure}
\begin{subfigure}[t]{0.33\linewidth}
    \includegraphics[width=\linewidth]{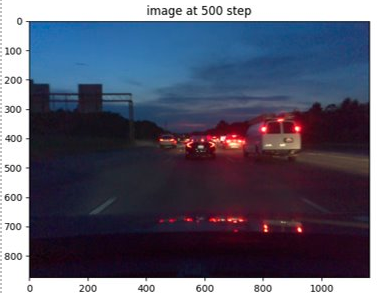}
    \caption{Front view at step 500 (5s)}
\end{subfigure}%
\begin{subfigure}[t]{0.33\linewidth}
    \includegraphics[width=\linewidth]{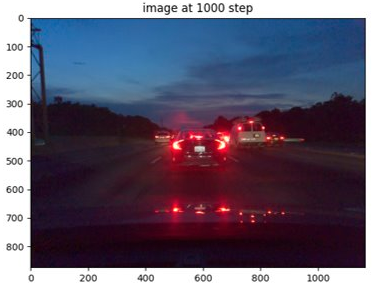}
    \caption{Front view at step 1000 (10s)}
\end{subfigure}
\\
\begin{subfigure}[t]{0.54\linewidth}
    \includegraphics[width=\linewidth]{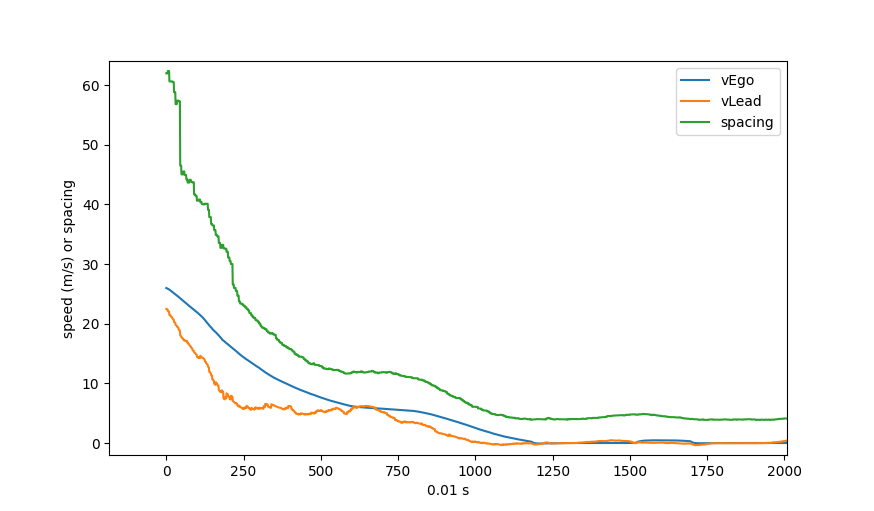}
    \caption{Speed and spacing profile}
\end{subfigure}
\begin{subfigure}[t]{0.42\linewidth}
    \includegraphics[width=\linewidth]{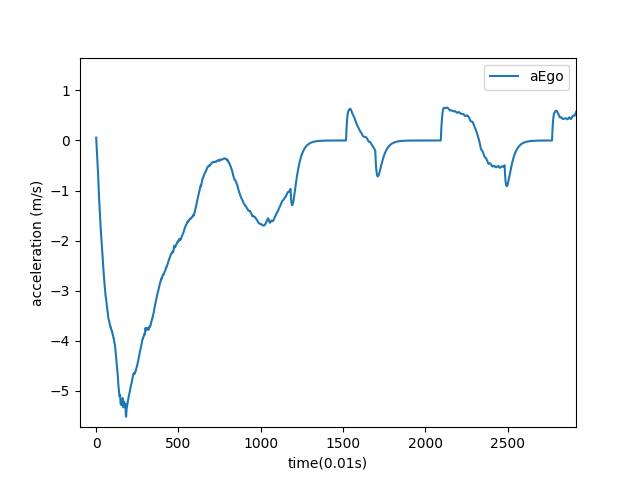}
    \caption{True acceleration of the vehicle}
\end{subfigure}
\caption{An example of the abrupt large brake on highway: the incident happened when a congestion wave hit back very fast unexpectedly, forcing the driver in the ego vehicle to apply a very large brake to avoid a crash.}
\label{dangerous-screeshots}
\end{figure}

It is not uncommon to encounter a congestion wave in highway traffic that travels fast towards upstream and forces drivers to brake hard to avoid a collision. While human drivers can avoid crashes through hard brakes, ACC's reaction becomes very limited in comparison. The situation is even more dangerous when the lead vehicle quickly drops to a low speed or even a full stop because the radar is more likely to drop the frame when detecting low-speed or still objects. Many ACC user instruction has mentioned this notorious radar issue and only very recent car models can cruise below 25 mph. A more recent talk from Tesla revealed that radar may fail to detect slow-moving or still objects, which means losing a leader in the ACC, which explains the failure to brake or even accelerate instead. Occasionally the authors encountered and recorded one of those dangerous situations described above; see Fig.\ref{dangerous-screeshots}. Notice that both the lead and the follower vehicles dropped speeds from around 55 mph to zero in 10 seconds, where the maximum deceleration is around $-5m/s^2$, which is almost double the limit we saw in the ACC data. The incident did not lead to a crash thanks to the quick and sufficient reaction from the human driver. It is questionable whether a bounded ACC deceleration model can safely brake in this situation.

The response delay, restricted deceleration capability, and insufficient low-level control can lead to very short spacing when the leader reduces its speed abruptly, causing great safety concerns. From the author's experiences, some market ACCs are likely to disengage in those emergencies when the gap is quickly reducing, i.e. the time to collision is dropping very fast. We conjecture such design is to force the human driver to take over in situations that require large deceleration values out of the bounds in ACC design. However, one can imagine this deficiency in ACC is really concerning which needs more investigation.








\section{Discussions and conclusions}

Considering the impact of acceleration and deceleration limits, we argue that a linear ACC model in the literature should extend the parameter space to $\mathcal{S} = \{k, \tau, a_0, \beta, d_0, \theta \}$, which correspond to the controller sensitivity, the desired headway, and the set of parameters for acceleration/deceleration constraints. Now we use $\mathcal{L}_{acc} (k, \tau, a_0, \beta, d_0, \theta) $ to denote a linear ACC planner with a set of the parameters.

\begin{figure}[htbp]
\centering
\begin{subfigure}[t]{0.49\linewidth}
    \includegraphics[width=\linewidth]{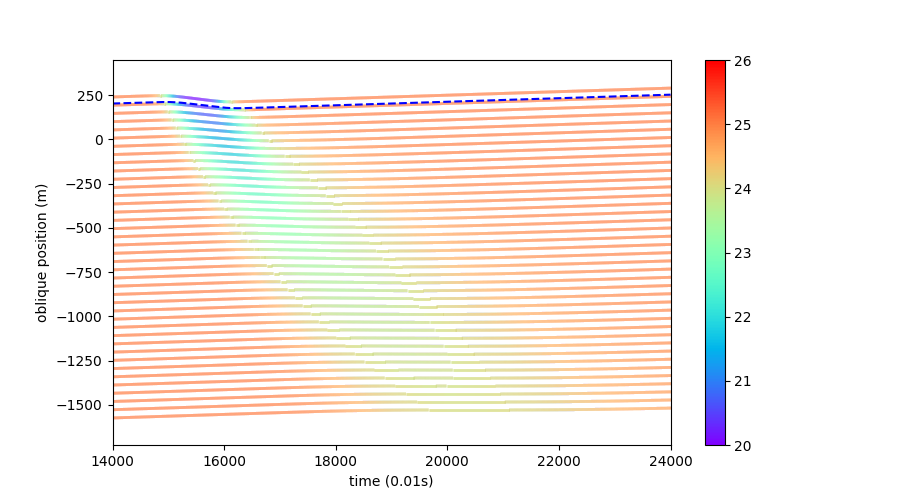}
    \caption{A string stable ACC platoon}
\end{subfigure}%
\begin{subfigure}[t]{0.49\linewidth}
    \includegraphics[width=\linewidth]{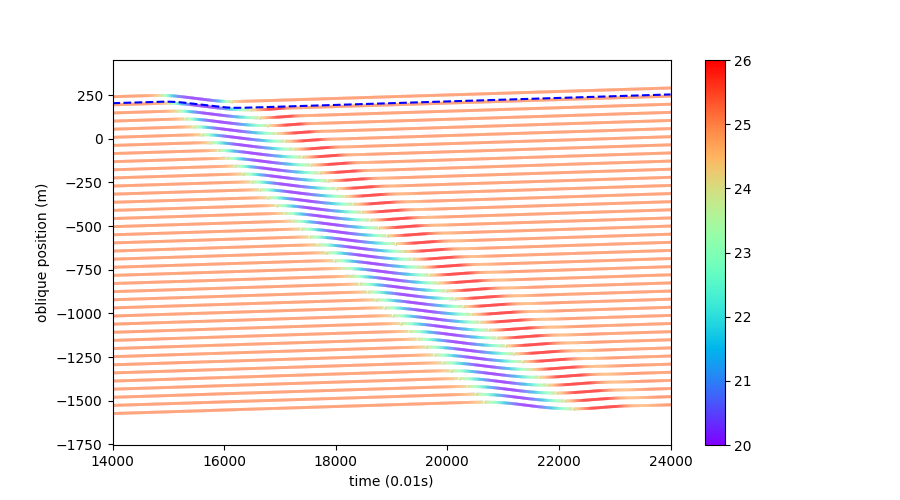}
    \caption{A marginal string stable ACC platoon}
\end{subfigure}
\caption{Comparing the ACC platoon consisting of string stable and marginal string stable models}
\label{margional-ss}
\end{figure}

The paper derived the condition for SS and also the marginal SS. While being stable indicates a smaller sensitivity factor $k$ for the linear factory ACC, the safety concern suggests a large $k$ is necessary. To balance, we suggest the marginal SS over SS for ACC platoons. The comparison results are shown in Fig.\ref{margional-ss}. In (b), the marginal SS does not amplify the speed perturbation and more importantly, it leads to shorter queue length compared to the string stable models in (a), which has a wide tail of congestion area.  

\begin{figure}[htbp]
\centering
\includegraphics[width=0.7\linewidth]{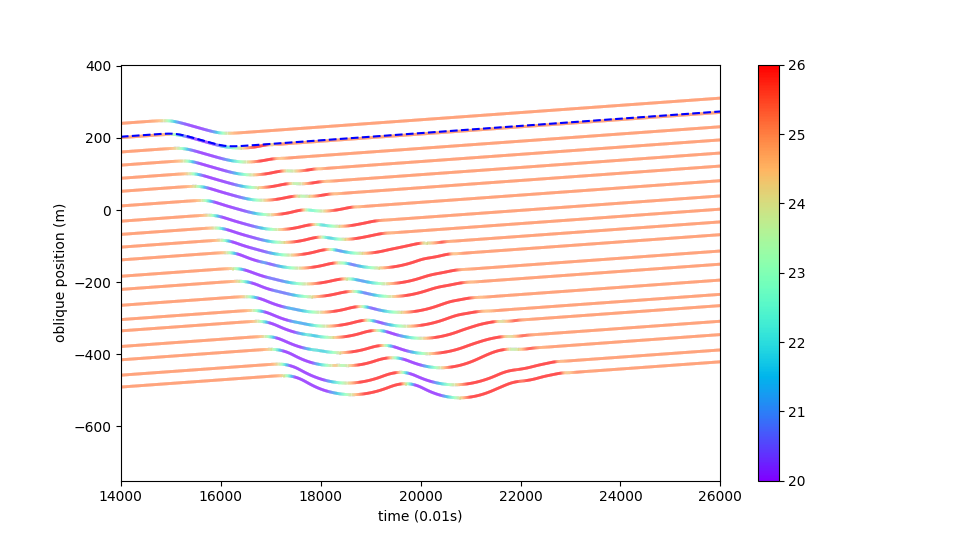}
\caption{Mixed ACC platoon with different parameters}
\label{random-platoon}
\end{figure}

In real-world traffic, the marginal SS cannot be guaranteed, and most likely we will have different ACC parameters which can be randomly sampled from some distribution. The impact of the string stability and the acceleration/deceleration bounds in a mixed platoon are certainly worth more investigation.
For example, the authors show a randomized heterogeneous ACC platoon where the lead vehicle creates a speed perturbation. The reactions of the mixed platoon are reported in Fig.\ref{random-platoon}, where we see multiple waves upstream of the platoon, which can be traced back to the overshoots caused by acceleration/deceleration bounds as shown in Section 3. 

\begin{figure}[htbp]
\centering
\includegraphics[width=0.6\linewidth]{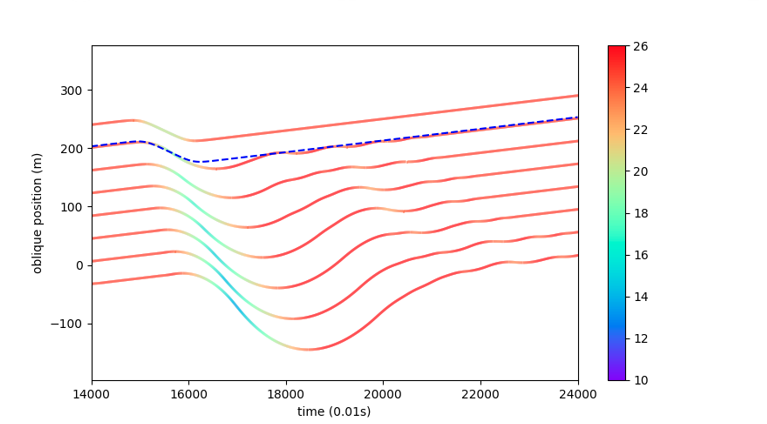}
\caption{Impact of acceleration/deceleration limits in string unstable platoons}
\label{safety-platoon}
\end{figure}

The paper also shared our concern about the deceleration limits in ACC design. We argue that combined with string unstable CF models,  the impact of acceleration/deceleration limits can be more concerning, which may lead to unsafe spacings or even crashes. Using the mixed platoon experiments, we show the safety hazard in Fig.\ref{safety-platoon}. More investigation is certainly needed to reduce those safety hazards in mixed ACC traffic. 

The paper pointed out the acceleration/deceleration limits affect the SS by first creating extra large/small gaps then the follower needs to overshoot for compensation. Such finding also sheds some new insights to better understanding human behaviors. We conjecture that the string instability of human driving may largely result from the reaction delay. Suppose the lead vehicle unexpectedly changes its speed, the reaction delay of the following driver may cause the spacing between the two vehicles to increase/decrease to a surprisingly large/small value when the leader accelerates/decelerates. As a result, the following driver has to hard press the gas/brake in compensation for the extra/insufficient spacing. Similar to the impact of acceleration limit, the follower usually needs to overshoot the lead speed to maintain the gap. 

\section*{Acknowledgment}
This research was funded by NSF Awards \#1932451 and \#1826162.

\appendix

\section{ODE solution of the factory linear ACC}
The transfer function obtained above indicates the ratio between the output (follower trajectory) and the input (leader) in the s plane. To provide better intuition, we examine the input-output relationship using a time-domain ODE approach by assuming a general perturbation of a sine wave, where the lead speed is as follows: 

\begin{equation}
    v_{lead}(t) = v_{eq} + M \sin(\omega t) \quad \text{for \quad} t \in [0,w]
    \label{lead-speed-t}
\end{equation}
where $M$ and $\omega$ are the amplitude and frequency of the signal respectively.

For the factory linear ACC model:
\begin{align}
    v_{ego}(t+\Delta t) &= v_{lead}(t) + k_v [x_{lead}(t)-x_{ego}(t)-\tau v_{lead}(t)-\delta]\\
    a_{ego} &= a_{lead} (1-k_v \tau) + k_v (v_{lead}-v_{ego})
    \label{factory-ACC-accel-t}
\end{align}

Substitute \eqref{lead-speed-t} into \eqref{factory-ACC-accel-t}, we have:
\begin{align}
      a_{ego}(t) &= k_v (v_{eq} + M \sin(wt)-v_{ego}(t))+(1-k_v \tau) M w \cos(wt) \\
      \dot{v}_{ego}(t) & = k_v v_{eq} + k_v M \sin(wt)+ (1-k_v \tau) Mw \cos(wt) -k_v v_{ego}(t) \label{ODE}
\end{align}

Solving the differential equation in \eqref{ODE}, we can obtain:
\begin{equation}
    v_{ego}(t) = \frac{k_v^2 v_{eq} +e^{-k_v t} k_v^2 M \tau \omega+ v_{eq} \omega^2 -k_v^2 M \tau \omega \cos(t \omega)  + M(k_v^2-k_v \tau \omega^2 +\omega^2)\sin(t \omega)}{k_v^2 + \omega^2}
    \label{solution-ode-factory}
\end{equation}

\begin{figure}
    \centering
    \includegraphics[width = 0.5\textwidth]{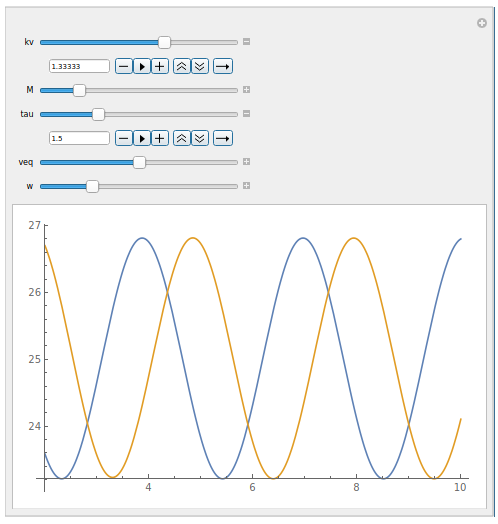}
    \caption{ODE solution to the linear factory ACC model: when $\tau=2/k_v$ holds, the marginal SS of the model does not depend on all the other variables including $M, \omega, v_{eq}$}
    \label{ode_manipulate}
\end{figure}
\FloatBarrier

For validation, we plot the leader trajectory \eqref{lead-speed-t} and the ODE solution of the follower trajectory \eqref{solution-ode-factory} in Fig.\ref{ode_manipulate}. Through animation one can see the SS does not depend on the signal magnitude $M$, the frequency $\omega$ and the origin speed level $v_{eq}$ if no acceleration limit is considered. For example, the marginal SS holds as long as $k_v = 2/\tau$.

Note that the ODE solution indicates the response time is equal to $\tau$.

\section{Solve the overshoot speed analytically}

The maximum spacing at $T_1$ is:
\begin{align}
    s(T_1) &= \int_{T_0}^{T_1} (v_{lead}(t) - v_{ego}(t)) dt \\
    & = \int_{T_0}^{T_1} v_{lead}(t) dt -\int_{T_0}^{T^1} [(v_0-v_c-a_0/\beta)e^{-\beta t} + a_0/\beta + v_c] dt \\
    &  = \int_{T_0}^{T_1} v_{lead}(t) dt + \frac{\beta v_{lead}(T_1)  - \beta v_c- a_0}{\beta^2}   - \frac{(\beta v_0-\beta v_c-a_0)}{\beta^2} e^{-\beta T_0} - (a_0/\beta + v_c) (T_1-T_0)
    \label{max_spacing}
\end{align}

From \eqref{max_speed_traj}, $T_1$ can be solved:
\begin{align}
     T_1 = -1/\beta \cdot \ln(\frac{v_{lead}(T_1) - a_0/\beta - v_c}{v_0-v_c-a_0/\beta}) \label{T1-time} 
\end{align}

Substituting $T_1$ into \eqref{max_spacing}, we can get: 
\begin{align}
    s(T_1) = \int_{T_0}^{T_1} v_{lead}(t) dt + \frac{\beta v_{lead}(T_1)  - \beta v_c- a_0}{\beta^2}   - \frac{(\beta v_0-\beta v_c-a_0)}{\beta^2} e^{-\beta T_0}  \\+ \frac{a_0 + \beta v_c}{\beta^2} \ln(\frac{v_{lead}(T_1) - a_0/\beta - v_c}{v_0-v_c-a_0/\beta}) + (a_0/\beta + v_c) T_0
    \label{simplifed-max_spacing}
\end{align}

From $T_1$, the $v_{pid}$ increases at the rate of the maximum acceleration in \eqref{accel_bound}, and the $v_{target}$ decreases with the rate of $k \dot{s} = k v_{ego}$. The maximum speed of $v_{ego}$ approximately happens at the time $v_{pid}$ and $v_{target}$ crosses, if we ignore the small impact of integral overshoot. 

At $T_1$, $v_{target}(T_1) = v_{lead}(T_1) + k (s(T_1) - \tau v_{lead}(T_1) -\delta)$, and $v_{pid}(T_1) = v_{lead}(T_1)$. Assuming the acceleration limit follows the model in \eqref{accel_bound}, then we obtain the time when the $v_{pid}$ and $v_{target}$ crosses:
\begin{align}
    \int_{T_1}^{T_4} [a_{ego}(t) + k v_{ego}(t)]dt & = v_{target}(T_1) - v_{pid}(T_1) \\
    \int_{T_1}^{T_4} [a^*(t) + k v^*(t)]dt & = k [s(T_1) - \tau v_{lead}(T_1) -\delta] 
\end{align}

For simplicity, assume a constant $a^*$ from $T_1$ to $T_4$, which leads to: 
\begin{align}
    \int_{T_1}^{T_4} [a^* + k v^*(t)]dt & \approx k [s(T_1) - \tau v_{lead}(T_1) -\delta] \\
    \int_{T_1}^{T_4} a^* dt + \int_{T_1}^{T_4} k v^*(t)]dt & \approx k [s(T_1) - \tau v_{lead}(T_1) -\delta] \\
    a^* (T_4-T_1) + k (T_4-T_1) [v_{lead}(T_1)+a^*(T_4-T_1)/2] & \approx k [s(T_1) - \tau v_{lead}(T_1) -\delta] \\
    \frac{1}{2} a^*k \Delta T^2 + [k\cdot v{lead}(T_1)]\Delta T & \approx k [s(T_1) - \tau v_{lead}(T_1) -\delta] \label{solve_T4}
\end{align}
where $\Delta T = T_4 - T_1$, which can be derived by solving the quadratic equation in \eqref{solve_T4}.

Accordingly, the maximum overshoot speed $v_{os}$ is:
\begin{align}
    v_{os} = v_{lead}(T_1) + \Delta T a* \approx v_{lead}(T_1) + \Delta T (a_0 + [v_c - v_{lead}(T_1)] \beta )
\end{align}

The impact of the acceleration limit on speed overshooting is thus directly measured by solving $v_{os}$.

\bibliography{Reference}

\appendix

\nolinenumbers
\end{document}